\documentclass[english,
12 pt
]{article}
\usepackage[T1]{fontenc}
\usepackage[latin9]{inputenc}
\usepackage{graphicx}
\usepackage{babel}
\usepackage{amsmath}
\usepackage{color}

\makeatletter

\newcommand{\bsubs}{\begin{subequations}}
\newcommand{\esubs}{\end{subequations}}

\newcommand{\be}{\begin{equation}}
\newcommand{\ee}{\end{equation}}

\newcommand{\bea}{\begin{eqnarray}}
\newcommand{\eea}{\end{eqnarray}}

\begin{document}

\title{Kenneth Geddes Wilson \\(1936-2013)\\Physicist who changed both the substance and style of theoretical science\\
 }
\author{ Leo P. Kadanoff\footnote{e-mail:  leop@UChicago.edu} \\
\\
The Perimeter Institute,\\
Waterloo, Ontario, Canada \\
\\
and \\
\\
The James Franck Institute\\
The University of Chicago
\\ Chicago, IL USA}

\maketitle

\newpage{}
Before Kenneth Wilson's work,  calculations  in particle physics were plagued ed by infinities.   Results came from  a workaround called {\em renormalization}, described by one of its inventors,  P.A.M. Dirac, in quite unflattering terms:  "I might have thought that the new ideas were correct if they had not been so ugly."   However, in  the 1970s, Wilson reformulated this method to eliminate its {\em ad hoc} elements.  Almost immediately, renormalization became a respectable and widely used tool, forming the basis of literally thousands of papers in condensed matter and particle physics.  It is now our primary method for seeing the connection among different theories in the physical sciences.

Wilson himself started life in the midst of theoretical science.  His grandfather taught at MIT; his father was a theoretical chemist at Harvard.   Ken's thesis adviser, Murray Gell-Mann of Cal Tech, was and is  a particle theorist of exceptional depth and breadth.  Like his father, Ken was appointed to the Society of Fellows at Harvard, a group of young scholars picked for their exceptional promise and then given no responsibilities whatsoever. 

Appointed to the  junior  faculty at Cornell in 1963, Wilson proved a disappointment, at least to those  who vet promotions by counting the number and impact of publications.  Despite having no journal publications before 1969, Wilson's exceptional promise,brought him tenure after  three years and full professorship after three more.  These promotions proved prescient since, in 1971, Wilson revolutionized the mathematical sciences by reconstructing  the renormalization method.

Both particle and condensed matter physics calculate results for length scales set by our experimental apparatus in terms of processes occurring at much shorter distances.  However, observation-based theories, whether they be of the elasticity of materials or of the collisions of observable ``elementary'' particles, cannot accurately describe natural forces acting at small scales.  
 This weakness showed up in midcentury particle physics via the generation of infinities in most calculations.  Before Wilson's work, the renormalization method achieved finite results by replacing the infinite quantities within the theory by empirically derived, finite quantities.

During the 1960s, Wilson had been thinking deeply about   particle physics while noting  the close analogy of this area  with  condensed matter physics, particularly  with the study of phase transitions. (A phase transition is the change of matter from one form to another, for example the boiling of liquid water to produce a vapor.) Phase transitions had a rich experimental, analytical and numerical tradition,  based upon well-known science at the atomic or molecular level. As a result,  in this area theory could be verified or disproven by comparison with known results.

In the analysis of both particles and condensed matter, the same physical system might be described by relating the maths of  different observers each focusing upon processes at a different scales of length. 
Wilson added two new elements to this old idea of changing length scales. First he broadened the  calculation of the connection between the different scales to include all physical processes, not just the few processes that would be likely to show infinities.    The second was to posit a final result for changes in length scale, called a ``fixed point,"  at which the system would  become scale-independent.  This fixed point could then serve as the basis of a deep analysis of the physical situation.   

Soon after Wilson first  described his new ideas, he worked with Michael Fisher  to calculate the primary facts about the liquid-gas phase transition.  A amazingly close agreement with experiment helped accelerate the acceptance of the new theory. 

In this new form, renormalization has proven to be substantially more than a technical tool.  It  connects the behavior at different scales.  For example, it builds a bridge between the behavior of molecules and the observed macroscopic properties of materials.  The connection among "laws of nature" in different regimes of energy, length, or aggregation is the root subject of physics. Thus Wilson has provided us with the single most relevant tool for understanding physics.  

That is Ken's main accomplishment.  But it is not all.  He has helped add to our profession a new style of work and of thinking.  When he said that we should do renormalization by looking at {\em all} the processes that might arise in changing  in length scales, he was asking for something impossible.  Nobody can keep track of hundreds of different processes.  But a computer can do so....     Wilson was implicitly suggesting we might develop new work-styles and new scientific areas by emulating computers and computer-programs.

Ken's engagement in computer-use led to  his sponsorship of proposals for supercomputers and supercomputer centers, and then morphed into the design of computer programs to permit flexible use of very large computers.  This work was, in part, carried out with Ken's wife, the computer scientist Alison Brown.  
 
Ken was unfailingly generous  to those working in his area.  One evidence of generosity is the very careful credit given to previous workers  in Ken's papers. For example, I note  Ken's  kind (and almost unprecedented) inclusion of my name in the title(!) of his first, great, renormalization paper.   In addition, I remember  a trip to Cornell which I took to learn about fermions  in a paper I was writing about quarks and strings.  My visit elicited a Wilson tutorial on Grassman variables, which I then used in my paper.  

The brilliance of Kenneth G. Wilson was dazzling, but he never tried to outshine those about him.   He was all quiet competence and deep accomplishments.

\section{Acknowledgments} This work was was supported in part by the University of Chicago MRSEC program under NSF DMR-MRSEC  grant number 0820054.  It was also supported in part by the Perimeter Institute,  which is supported by the Government of Canada through Industry Canada and by the Province of Ontario through the Ministry of Research and Innovation.

\end{document}